\documentclass[fleqn,twoside]{article}
\usepackage{mod_espcrc2}

\usepackage{graphicx}
\usepackage[figuresright]{rotating}

\newcommand{\AmS}{{\protect\the\textfont2
  A\kern-.1667em\lower.5ex\hbox{M}\kern-.125emS}}
\newcommand{\beq}{\begin{equation}}
\newcommand{\eeq}{\end{equation}}
\newcommand{\bea}{\begin{eqnarray}}
\newcommand{\eea}{\end{eqnarray}}
\newcommand{\rmR}{{\rm R}}

\def\za{Z_{\rm A}}
\def\zv{Z_{\rm V}}
\def\zp{Z_{\rm P}}

\def\zg{Z_{\rm g}}
\def\zm{Z_{\rm m}}
\def\zmu{Z_{\mu}}

\def\wt{\widetilde}

\def\bfx{{\bf x}}

\def\half {\frac{1}{2}}

\def\proof{\noindent{\sl Proof:}\kern0.6em}

\def\frac#1#2{\hbox{$#1\over#2$}}
\def\dual{\mathstrut^*\kern-0.1em}

\def\lvec#1{\setbox0=\hbox{$#1$}
    \setbox1=\hbox{$\scriptstyle\leftarrow$}
    #1\kern-\wd0\smash{
    \raise\ht0\hbox{$\raise1pt\hbox{$\scriptstyle\leftarrow$}$}}
    \kern-\wd1\kern\wd0}
\def\rvec#1{\setbox0=\hbox{$#1$}
    \setbox1=\hbox{$\scriptstyle\rightarrow$}
    #1\kern-\wd0\smash{
    \raise\ht0\hbox{$\raise1pt\hbox{$\scriptstyle\rightarrow$}$}}
    \kern-\wd1\kern\wd0}

\def\nabstar#1{\nabla\kern-0.5pt\smash{\raise 4.5pt\hbox{$\ast$}}
               \kern-4.5pt_{#1}}

\def\drvstar#1{\partial\kern-0.5pt\smash{\raise 4.5pt\hbox{$\ast$}}
               \kern-5.0pt_{#1}}

\def\momp#1#2{
    \setbox0=\hbox{${#1}$}\setbox1=\hbox{${#1}_{#2}$}
    {#1}_{#2}\kern-\wd1\kern\wd0
    \smash{\raise4.5pt\hbox{$\scriptscriptstyle +$}}}
\def\momm#1#2{
    \setbox0=\hbox{${#1}$}\setbox1=\hbox{${#1}_{#2}$}
    {#1}_{#2}\kern-\wd1\kern\wd0
    \smash{\raise4.5pt\hbox{$\scriptscriptstyle -$}}}
\def\mompm#1#2{
    \setbox0=\hbox{${#1}$}\setbox1=\hbox{${#1}_{#2}$}
    {#1}_{#2}\kern-\wd1\kern\wd0
    \smash{\raise4.5pt\hbox{$\scriptscriptstyle \pm$}}}
\def\smomp#1#2{
    \setbox0=\hbox{${#1}$}\setbox1=\hbox{${#1}_{#2}$}
    {#1}_{#2}\kern-\wd1\kern\wd0
    \smash{\raise3pt\hbox{$\scriptscriptstyle +$}}}
\def\smomm#1#2{
    \setbox0=\hbox{${#1}$}\setbox1=\hbox{${#1}_{#2}$}
    {#1}_{#2}\kern-\wd1\kern\wd0
    \smash{\raise3pt\hbox{$\scriptscriptstyle -$}}}
\def\smompm#1#2{
    \setbox0=\hbox{${#1}$}\setbox1=\hbox{${#1}_{#2}$}
    {#1}_{#2}\kern-\wd1\kern\wd0
    \smash{\raise3pt\hbox{$\scriptscriptstyle \pm$}}}
\def\si{\kern1pt{\rm si}}
\def\co{\kern1pt{\rm co}}

\def\Nf{N_{\rm f}}
\def\qbar{\bar{q}}
\def\psibar{\bar{\psi}}

\def\psiprime{\psi\kern1pt'}
\def\psibarprime{\psibar\kern1pt'}
\def\rhoprime{\rho\kern1pt'}
\def\rhobar{\bar{\rho}}
\def\rhobarprime{\rhobar\kern1pt'}
\def\rhobartilde{\kern2pt\tilde{\kern-2pt\rhobar}}
\def\rhobartildeprime{\kern2pt\tilde{\kern-2pt\rhobar}\kern1pt'}

\def\zetabar{\bar{\zeta}}
\def\zetaprime{\zeta\kern1pt'}
\def\zetabarprime{\zetabar\kern1pt'}
\def\zetar{\zeta_{\raise-1pt\hbox{\sixrm R}}}
\def\zetabarr{\zetabar_{\raise-1pt\hbox{\sixrm R}}}

\def\phiimpr{\phi_{\kern0.5pt\hbox{\sixrm I}}}

\def\diracstar#1#2{
    \setbox0=\hbox{$\gamma$}\setbox1=\hbox{$\gamma_{#1}$}
    \gamma_{#1}\kern-\wd1\kern\wd0
    \smash{\raise4.5pt\hbox{$\scriptstyle#2$}}}

\def\ba{b_{\rm A}}
\def\tba{\tilde{b}_{\rm A}}
\def\bp{b_{\rm P}}

\def\bg{b_{\rm g}}
\def\bm{b_{\rm m}}
\def\tbm{\tilde{b}_{\rm m}}
\def\bmu{b_{\mu}}

\def\ca{c_{\rm A}}

\def\csw{c_{\rm sw}}

\def\f1{f_1}

\def\h1{h_1}

\def\opprime#1{\setbox0=\hbox{${\cal O}$}\setbox1=\hbox{${\cal O}_{\rm #1}$}
    {\cal O}_{\rm #1}\kern-\wd1\kern\wd0
    \smash{\raise4.5pt\hbox{\kern1pt$\scriptstyle\prime$}}\kern1pt}

\def\ophatprime#1{\setbox0=\hbox{$\widehat{\cal O}$}
    \setbox1=\hbox{$\widehat{\cal O}_{\rm #1}$}
    \widehat{\cal O}_{\rm #1}\kern-\wd1\kern\wd0
    \smash{\raise4.5pt\hbox{\kern1pt$\scriptstyle\prime$}}\kern1pt}

\def\bopprime#1{\setbox0=\hbox{${\cal O}$}\setbox1=\hbox{${\cal O}_{\rm #1}$}
    {\cal L}_{\rm #1}\kern-\wd1\kern\wd0
    \smash{\raise4.5pt\hbox{\kern1pt$\scriptstyle\prime$}}\kern1pt}

\def\blagprime#1{\setbox0=\hbox{${\cal B}$}\setbox1=\hbox{${\cal B}_{#1}$}
    {\cal B}_{#1}\kern-\wd1\kern\wd0
    \smash{\raise5.2pt\hbox{\kern1pt$\scriptstyle\prime$}}\kern1pt}

\def\muq{\mu_{\rm q}}
\def\mq{m_{\rm q}}

\def\mc{m_{\rm c}}

\def\za{Z_{\rm A}}
\def\zv{Z_{\rm V}}
\def\zp{Z_{\rm P}}

\def\zg{Z_{\rm g}}
\def\zm{Z_{\rm m}}
\def\zmu{Z_{\mu}}

\def\msbar{{\rm \overline{MS\kern-0.05em}\kern0.05em}}

\def\tmW{{\rm tmW }}

\newcommand{\bes}{\begin{eqnarray}}
\newcommand{\ees}{\end{eqnarray}}

\title{Wilson fermions with chirally twisted mass\thanks{Based on a talk given at
Lattice 2002, the
XX International Symposium on Lattice Field Theory, held on June, 24--29, 2002
at M.I.T. Cambridge, Massachusetts, U.S.A.} 
      }

\author{R. Frezzotti\address[infnmi]{I.N.F.N. Milano and University 
        of Milano Bicocca, 
        Piazza della Scienza 3, I-20126 Milano, Italy}
        }
       
\begin{document}

\begin{abstract}
 Lattice formulations of QCD with Wilson fermions and a chirally
twisted quark mass matrix provide an attractive framework for 
non-perturbative numerical studies. Owing to reparameterization
invariance, the limiting continuum theory is just QCD. 
No spurious quark zero modes, which are responsible for the problem
with exceptional configurations, can occur at finite values of
the quark mass. 
Moreover, the details of the lattice formulation can be adjusted 
so as to simplify the renormalization and the O($a$) improvement of
several quantities of phenomenological relevance. The first exploratory
studies in the quenched approximation yield very encouraging results.
\vspace{1pc}
\end{abstract}

\maketitle
 
\section{Introduction}

In recent years a lot of progress has been achieved about
lattice regularizations of gauge theories with fermions~\cite{Kiku_rev}. 
On one hand, local 
(but non--ultralocal) fermionic actions that enjoy 
a lattice form of chiral symmetry and (almost) ideal renormalization
properties have been discovered and put at work~\cite{Giusti_rev}:
for all of them the critical Dirac operator satisfies
the celebrated Ginsparg-Wilson relation~\cite{GW}.
On the other hand, more traditional formulations of lattice QCD (LQCD) 
based on Wilson and staggered quarks have been refined and widely used in
realistic computations of hadronic observables and matrix elements.
The simplest of these computations are currently performed
with dynamical quarks.

In this contribution, I report on recent developments about
Wilson fermions, which remove practical obstructions in dealing
with light quarks and simplify the renormalization and O($a$) 
improvement of a number of observables.
Improvements in the formulations of LQCD with
staggered quarks are discussed elsewhere~\cite{AnnaH_rev}.

After shortly recalling the status of LQCD with Wilson quarks and its problem
with exceptional configurations (sect.$\,2$), I introduce --for
the case of $\Nf=2$ flavours-- the formulation
with chirally twisted mass (sect.$\,3$). Then I
discuss its basic properties (sect.$\,4$), 
the first non-perturbative studies (sect.$\,5$) and
analogous lattice formulations of QCD with $\Nf>2$ quark flavours,
which can simplify the renormalization of some operators
of the effective weak Hamiltonian (sect.~$\,6$). 

\section{Wilson fermions}

The well known lattice formulation introduced by Wilson~\cite{WILS74} 
provides a gauge invariant regularization for QCD with any number
of quark flavours: the action is ultralocal and respects
the global flavour symmetries of QCD, but all axial symmetries are
broken by the Wilson term. This is no principle problem, as
the flavour chiral invariance $SU(\Nf) \otimes SU(\Nf)$ can
be restored in the (quantum) continuum limit~\cite{chiral_recov}, while the axial
$U(1)$ invariance of the classical continuum theory is broken
by quantum fluctuations. 

The lack of chiral symmetry entails however some practically
important consequences. First, complicated patterns
of operator mixings arise, so that in many cases several operator
subtractions\footnote{The corresponding coefficients can be computed
in perturbation theory and beyond,
e.g. by requiring the chiral Ward identities to hold up to
cutoff effects~\cite{chiral_recov}.} are needed
in order to restore the chiral multiplet structure.
Then, the leading deviations 
of renormalized quantities from their continuum limit values are
linear in the lattice spacing $a$ and typically non-negligible.
Last but not least, the absence of a lower bound
on the norm of the eigenvalues of the massive Dirac 
matrix may lead to unphysical divergences in 
fermionic observables on non-trivial gauge backgrounds.

\subsection{O($a$) improved Wilson fermions}

The problem with the leading cutoff effects has found a clean solution
via Symanzik's improvement programme~\cite{SYM,SW,HMPRS,LUS1,LSSWW},
which allows to define and compute on-shell renormalized
quantities with leading cutoff effects of order $a^2$,
though at the price of tuning the coefficients of further counterterms\footnote{
Implementing this programme may be non-trivial
and CPU-time demanding, e.g. in the theory with non-degenerate and/or 
dynamical quarks, as well as for operators with complicated mixings~\cite{impr_review}.}.
The LQCD action for the O($a$) improved theory with
$\Nf$ quark flavours reads:
\beq \label{Wils_action}
S_{\rm g}[U;g_0^2] + a^4 \sum_x \psibar(x) 
\left[(D_{\rm W}^c [U] + M_{\rm q}) \psi\right](x)
\eeq 
where $S_{\rm g}[U;g_0^2]$ is the pure gauge action with coupling 
$g_0^2$, $M_{\rm q} = {\rm diag}(\mq^u, \mq^d, \mq^s, \dots ) $ is the 
subtracted ($M_{\rm q} = M_0 - \mc$) quark mass matrix and
\beq \label{Wils_Dirac_op}
D_{\rm W}^c [U] =
\left\{  \gamma \widetilde\nabla
         - \frac{a}{2} \nabla^\star \nabla
        + \csw \frac{a}{4} i \sigma \hat{F} \right\} [U]
        + \mc \; 
\eeq
represents the critical Wilson--Dirac operator.
Omitting Lorentz indices, we denote by $\nabla$ ($\nabla^\star $, 
$\widetilde\nabla$) the forward (backward, symmetrized) covariant 
lattice derivative and by $\hat{F}_{\mu\nu}$ the clover lattice
discretization of $F_{\mu\nu}$. The coefficients $\csw$ and $a\mc$ 
are dimensionless functions\footnote{At the non--perturbative level, 
$\csw$ and $a\mc$ are uniquely defined only up to O($a$) and O($a^3$)
corrections~\cite{LSSWW}, respectively.} 
of $g_0^2$ and $\Nf$. 

\subsection{Exceptional configurations}

In quenched simulations of LQCD 
on relatively coarse lattices 
and for {\em moderately light} quark flavours,
gauge configurations can be sampled
where the Wilson--Dirac matrix
has one or few eigenvalues with norm {\em exceptionally close to zero},
i.e. much smaller than on the other configurations. 
The corresponding eigenvectors of the Wilson--Dirac matrix are
referred to as ``spurious'' quark zero modes, because in a 
chiral invariant formulation of QCD the Dirac matrix
can have zero modes only if some quark flavour is massless.  

On gauge configurations with spurious quark zero modes
fermionic observables may undergo fluctuations that exceed in modulus
the typical ones by orders of magnitudes: in this sense,  
these configurations appear to be ``exceptional''. 
Moreover, increasing the statistics does not reduce in general
the standard deviation of the observables, as further, even larger
fluctuations may occur. A reliable statistical analysis of the
fermionic observable becomes hence impossible~\cite{LSSWW,BDEHT}.
When employing the non-perturbatively O($a$) 
improved~\cite{LSSWW} Wilson action, eq.~(\ref{Wils_action}), and the plaquette gauge action,
this problem is felt at values of $\mq$ of about one half the value
of the strange quark's mass if $a \simeq 0.1$~fm~\cite{garden}.
With the unimproved Wilson action ($\csw=0$) quenched computations can be pushed
to somewhat lower values of the quark mass, but cutoff effects are harder
to control~\cite{CPPACS_wils}.

In unquenched simulations, the importance sampling
must give spurious quark zero modes a vanishingly small probability.
State-of-the-art algorithms, however, account for the
effects of sea quarks in a stochastic way: it is hence possible
that configurations with nearly zero-modes of the Dirac matrix are produced
in the updating process (e.g. a HMC trajectory). This would then result
in an exceptionally low acceptance rate\footnote{Exceptionally large values
of the product `driving force times MD integration step'' have been 
observed~\cite{instab_HMC} in simulations with the HMC algorithm.
The eigenvalues of the Dirac matrix were not simultaneously evaluated.},
which can possibly be avoided at the price of slowing down the algorithm. 
In the case of partially (un)quenched simulations, the problem
with spurious valence quark zero modes is only alleviated:
exceptional fluctuations of hadron correlators have indeed been
observed~\cite{UKQCD_pq}.
\begin{figure}[htb]
\vspace{-0pt}
\begin{center}
\includegraphics[angle=-90,width=12pc]{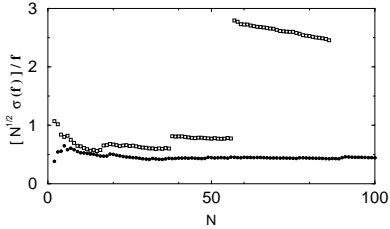}
\vspace{-35pt}
\end{center}
\caption{
Normalized standard deviation $\sqrt{N} \sigma(f)/f$
vs. the number of measurements $N$ in simulations
with standard (empty symbols) and chirally twisted
(filled symbols) Wilson quarks:
$f$ is a pion correlator at time separation $t=24a$.} 
\vspace{-10pt}
\label{FIG1}
\end{figure}

A typical example of exceptional configurations
is given in Fig.~\ref{FIG1}.
There we plot versus the number $N$ of independent measurements 
the relative standard deviation, multiplied by $N^{1/2}$, of a pion
channel correlator at fixed time separation: 
for the statistical analysis to be reliable, this
quantity should approach a constant value as $N \gg 1$. 
Our example refers to a {\em quenched} simulation at $\beta=6$ ($a \simeq 0.1$ fm)
on a $24^3 \cdot 48$ lattice, with degenerate quark masses such that
$m_{\rm PS}/m_{\rm V} \simeq 0.47$: employing the non-perturbatively
O($a$) improved action~(\ref{Wils_action}) leads to untolerably 
large and rare fluctuations of the observable (empty symbols).
On the other hand, one can work at equivalent parameters in a formulation
based on Wilson quarks with chirally twisted mass, see sect.$\;3$: of course, the 
bare quark mass parameters have to be adjusted so as to match the
value of the renormalized current quark mass ($m'_\rmR$ in the notation of sect.~4).
In this case the occurrence of spurious quark zero modes is avoided:
as shown in Fig.~\ref{FIG1} (filled symbols), there
are no exceptional fluctuations in the MC history of the pion
channel correlator, so that a reliable statistical result\footnote{The full statistics
was of $N=260$ measurements ~\cite{DFH_latt01buda}.}
can be obtained.

\section{Chirally twisted ``mass''}

The presence of chirally twisted ``mass'' terms is a general feature
of lattice gauge theories with Wilson fermions, as soon as the 
``physical'' fermion mass term is {\em not aligned} in chiral space 
to the Wilson term. Of course, the latter generates itself a ``mass'' 
term, whose divergent part requires appropriate subtraction.

\subsection{Twisted mass and parity}

To the best of my knowledge, Osterwalder and Seiler first 
proposed~\cite{Osterw_Seil} a lattice formulation where the 
fermionic action is of the form
\bea \label{tw_singlet_act}
S_{\rm f}^{(\phi)} = \sum_x \psibar'(x)
\left[ (D_{\phi}^c + \tilde{m}) \psi' \right](x)  \, ,
\nonumber \\
D_{\phi}^c \!=\!
\gamma \widetilde\nabla +  e^{-i \phi \gamma_5 }
[\tilde{m}_{\rm c} - \frac{a}{2} \nabla^\star \nabla
 + \tilde{c}_{\rm SW} \frac{a}{4} i \sigma \hat{F} ]
\, ,
\eea
namely with $\phi = \pi/2$ and $\tilde{c}_{\rm SW}=0$. 
The angle $\phi$ can be taken as
a measurement of the misalignment between the fermion mass term
($\propto \tilde{m}$) and the Wilson term, subtracted by a
critical mass and a clover-like term\footnote{In general,
$a\tilde{m}_{\rm c}$ and $\tilde{c}_{\rm SW}$ may differ
from $a\mc$ and $\csw$.}. 
It was also observed that lattice formulations with fermionic
action~(\ref{tw_singlet_act}) in general do not respect 
parity~\cite{Immirzi_et_al}. Indeed, as long as 
$\phi \neq 0 \;(\rm mod\; \pi)$,
quantum fluctuations generate in the effective action a term
$\propto {\rm tr} (F \widetilde{F})$, which survives to the 
continuum limit (if not canceled by a suitable counterterm).

Parity breaking is an unwanted feature if one is interested in QCD.  
One can show that, up to cutoff effects, parity breaking is avoided
by considering $\Nf >1$ quark flavours and giving the array
$\phi = [\phi^{(1)}, \dots , \phi^{(\Nf)}]$ a non-trivial flavour 
dependence such that $\sum_{f=1}^{\Nf} \phi^{(f)} = 0$.
In the case of $\Nf=2$, which is relevant e.g. for $u$ and $d$
quarks, one is then led to the
fermionic action:
\beq \label{tmQCD_act_pb}
S_{\rm f} = a^4 \sum_x
\qbar'(x) \left(
D_{\rm tmW}^c + \mq'
\right) q'(x)
\eeq
where $q'$ is a quark doublet field, the quark mass term ($\propto 
\mq'$) takes the usual form, but the critical Wilson-Dirac matrix is 
chirally twisted\footnote{The matrix $\tau^3$ in $D_\tmW^c$ acts in
flavour space.}:
\beq \label{tm_Wilsmatr}
D_\tmW^c \!=\! \gamma \widetilde\nabla + e^{-i \omega \gamma_5 \tau^3 }
  [\mc - \frac{a}{2} \nabla^\star \nabla
  + \csw \frac{a}{4} i\sigma \hat{F}]
\eeq
A (non-anomalous) change of quark field basis,
\beq \label{tm_rot}
q = e^{-i \omega \gamma_5 \tau^3 /2 } q' \, ,
\quad \quad \quad
\qbar = \qbar' e^{-i \omega \gamma_5 \tau^3 /2 } \, ,
\eeq
brings the action~(\ref{tmQCD_act_pb}) into the form 
\beq \label{tmQCD_act_ub}   
S_{\rm f} = a^4 \sum_x
\qbar(x) \left(
D_{\rm W}^c + \mq + i \muq \gamma_5 \tau^3
\right) q(x)
\eeq
where $\muq = \mq' \sin \omega$ is the twisted mass parameter
and $\mq = m_0 - \mc = \mq' \cos \omega$. 
The bare mass parameters are $\muq$
and $m_0$, as the $\mc$--dependence 
cancels between $\mq$ and $D_{\rm W}^c$.

The formulation with fermionic
action~(\ref{tmQCD_act_pb}) or~(\ref{tmQCD_act_ub})
is referred to as lattice twisted mass QCD (LtmQCD).
At the classical level, it obviously represents a
regularization of QCD with $\Nf = 2$ (mass--degenerate)
flavours. This property holds true at the quantum 
level~\cite{paper1}, see sect.$\,4$. 
%

\subsection{Protection against exceptionals }

The lattice theory with fermionic action~(\ref{tmQCD_act_ub}) has 
been studied (for $\csw=0$) to establish whether, at finite 
$g_0^2$ and for values of $am_0$ in a certain range,
parity and isospin can be spontaneously broken\footnote{In this
context $\muq$ played the role of an external zero--momentum source,
which can orient the vacuum in a certain direction (in flavour chiral
space), and the main interest was on the properties of the vacuum after
taking the thermodynamic and {\em zero--source} limits.}, as conjectured by
Aoki~\cite{aoki_phase}. It was noted~\cite{AG}
that the lattice Dirac matrix corresponding to eq.~(\ref{tmQCD_act_ub})
can not be singular on any gauge background as long as $\muq \neq 0$,
since
\bea \label{ag_bound}
0 \quad < \quad {\rm Det} [D_{\rm W}^c + \mq + i\muq \gamma_5 \tau^3 ] =
\nonumber \\
= {\rm det} [ (D_{\rm W}^c + \mq)^\dagger (D_{\rm W}^c + \mq) +
  \mu_{\rm q}^2 ]  \, .
\eea
Here ${\rm Det}[\dots]$ (${\rm det}[\dots]$) denotes the fermionic 
determinant in the two-flavour (one-flavour) space.


The one-flavour Dirac matrix corresponding to the 
action~(\ref{tw_singlet_act}) is also not singular on any
gauge configurations, as long as $\tilde{m} \sin \phi \neq 0$.
This is because $|{\rm det} [D_{\phi}^c + \tilde{m} ] |^2$ 
can be written in the same form as the r.h.s.\ of eq.~(\ref{ag_bound}) with 
$\mq \Rightarrow \tilde{m} \cos\phi$ and $\muq \Rightarrow \tilde{m} \sin\phi$.
Based on this property, the authors of Ref.~\cite{SCHI_et_al}
proposed to employ the fermionic lattice action~(\ref{tw_singlet_act})
with $\tilde{m}_{\rm c} = \mc$, $\tilde{c}_{\rm SW} = \csw$,
$\pi/2 \geq |\phi| > 0$ and $\tilde{m} > 0$ 
to avoid the occurrence of spurious quark zero modes in massive LQCD
with Wilson quarks. They presented numerical evidence, see Fig.~3 of
Ref~\cite{SCHI_et_al}, that exceptional configurations are avoided 
in the quenched approximation and discussed on a semiclassical level 
the relation between their lattice formulation and QCD in the continuum 
limit. It should be noted that
in the quenched model the parity breaking inherent to the
formulation with action~(\ref{tw_singlet_act}) reduces to a
mere O($a$) effect on renormalized quantities, but this is
no longer true with (partially) unquenched quarks.


\section {Basic properties of lattice tmQCD}

I discuss here the properties of LtmQCD as an ultraviolet regularization
of QCD with $\Nf=2$ mass-degenerate quark flavours\footnote{A satisfactory
regularization of $\Nf=2$ QCD with bare quark masses $\mq'
\pm \delta \mq'$ can be obtained  e.g. by adding to the action density
in eq.~(\ref{tmQCD_act_pb}) a term $\qbar'(x) \tau^1 \delta \mq' q'(x)$.}. 

\subsection{Symmetries and renormalizability}
The LtmQCD action, see eq.~(\ref{tmQCD_act_pb}) or~(\ref{tmQCD_act_ub}), 
is invariant under lattice gauge transformations and translations,
axis permutations, charge conjugation and the global
symmetry U(1) corresponding to conservation of the fermionic number. 
At $\muq \neq 0 \Leftrightarrow \omega = \arctan(\muq/\mq) \neq 0$, 
isospin symmetry is reduced to a U(1)--invariance with generator $\tau^3/2$
and axis reflections, such as parity, are no longer a symmetry. It is
important to note the residual invariance $P_{\rm F}$,
\bea \label{P_F_def}
U_0(x) \to U_0(x_P) \, ,  \;\quad
U_k(x) \to U_k^{-1}(x_P-a\hat{k}) \, , \!\!\!
\nonumber \\
q(x) \to \tau^1 \gamma_0 q(x_P) \, , \;\quad
\qbar(x) \to \qbar(x_P)  \gamma_0 \tau^1 \, , \;\quad
\eea
where $x_P = (x_0, -\bfx)$, as this symmetry rules out a term $\propto 
{\rm tr} (F \widetilde{F})$ in the quantum effective action.  
  
A standard analysis~\cite{paper1} based on lattice
symmetries and power counting shows that the model is renormalizable.
The relations between bare and renormalized action parameters take in
general the form (I recall $\mq = m_0 -\mc$):
\bea \label{reno_param_1}
g^2_\rmR & = & \zg(g_0^2,a\mq,a\muq;a\mu) \, g_0^2 \; , 
\nonumber \\
\mu_\rmR & = & Z_\mu(g_0^2,a\mq,a\muq;a\mu) \, \muq \; , 
\nonumber \\
m_\rmR & = & \zm(g_0^2,a\mq,a\muq;a\mu) \, \mq \; , 
\eea
but the renormalization factors $Z$ can be chosen to be independent
of $a\mq$ and $a\muq$ (mass independent schemes).
The additive renormalization of $m_0$ is independent of the quark mass
parameters~\cite{chiral_recov,testa98,paper2}: $\mc = \mc(g_0^2)$, up to 
intrinsic O($a$) corrections in the case of 
non-perturbative determinations of $\mc$.
The ratio $\zm/\zmu$ does not depend on the subtraction scale $\mu$,          
consistently with recovery of flavour chiral symmetry in the continuum limit.
Noting the bare lattice identity
\beq \label{exact_PCVC}
\partial^\star_\mu \wt{V}^b_\mu = -2 \muq \epsilon^{3bc} P^c \, ,
\eeq
where $\wt{V}^b_\mu$ denotes the one--point split vector current,
it is easy to argue~\cite{paper1} that $\zmu\zp=1$

\subsection{Continuum limit and cutoff effects}

Although the physical interpretation of the fermionic correlation functions
is most transparent in the quark basis corresponding to eq.~(\ref{tmQCD_act_pb}),
the renormalization of gauge--invariant correlation functions, including those 
with insertions of local operators, looks simpler in the quark basis corresponding
to eq.~(\ref{tmQCD_act_ub}). 
In this basis, indeed,  
the critical Wilson-Dirac matrix is given by $D_{\rm W}^c$ and the mixing
properties of the operators in the massless theory ($\muq = 0$, $m_0 = \mc$)
are usually well known: the construction of lattice fields that are multiplicatively 
renormalizable and respect, up to cutoff effects, the chiral multiplet structure
is hence straightforward. Concerning the overall, possibly scale--dependent
renormalization factors of the various operator multiplets, we assume for
simplicity~\cite{paper1} that they, as well as $\zg$, $\zm$ and $\zmu$, are chosen
to be independent of $\omega$.

In the quark basis of choice, the Ward identities of flavour chiral symmetry
read (for $b=1,2,3$):
\begin{eqnarray} \label{fch_WI}
\widetilde\partial_\mu ( A_\rmR)_\mu^b & \simeq & 2 m_\rmR ( P_\rmR)^b
+ i \mu_\rmR \delta^{3b} ( S_\rmR)^0  \nonumber \\
\widetilde\partial_\mu ( V_\rmR)_\mu^b & \simeq & 
- 2 \mu_\rmR \epsilon^{3bc} ( P_\rmR)^c \, ,
\end{eqnarray}
where the symbol $\simeq$ means equality up to O($a$) corrections and
renormalized operators are defined as usual with Wilson fermions\footnote{For
$b=3$ severe power-like divergences must be subtracted to define $( S_\rmR)^0$,
which is no problem in perturbation theory but is delicate at the non-perturbative level.}, e.g.
\beq \label{chir_curr}
( \, \{V_\rmR,A_\rmR\} \, )_\mu^b = 
\qbar \{\zv \gamma_\mu, \za \gamma_\mu \gamma_5 \} \half \tau^b q \, .
\eeq

The form of eq.~(\ref{fch_WI}) reminds us that appropriate linear
combinations of renormalized operators, reflecting the formal
change of variables~(\ref{tm_rot}),
should be taken, in order to obtain in the continuum limit operators
with well defined parity and isospin properties. 
Exploiting eq.~(\ref{fch_WI}) one can indeed extract $m_\rmR$ and $\mu_\rmR$ and
introduce the renormalized counterparts of $m'_{\rm q}$ and $\omega$:
\beq \label{pyth_ren_mass}
m'_\rmR = \sqrt{ m_\rmR^2 + \mu_\rmR^2 } \, ,
\quad\;
\alpha = {\rm arctan} \left( {\mu_\rmR \over m_\rmR}\right) \, .
\eeq
I remark that $\tan (\alpha ) = (\zmu / \zm) \tan(\omega)$
and $m'_\rmR = Z_{\rm m'}(g_0^2, \omega) \mq'$.
Then, if one considers (with $\bar{b}=1,2$) the renormalized operators 
\bea \label{primed_AVP}
( V'_\rmR )_\mu^3 & \equiv & ( V_\rmR )_\mu^3 \, ,
\quad\quad 
( A'_\rmR )_\mu^3 \;\; \equiv \;\; ( A_\rmR )_\mu^3 \, ,
\nonumber \\
( V'_\rmR )_\mu^{\bar{b}} &  \equiv &  
\cos(\alpha) ( V_\rmR )_\mu^{\bar{b}}
+ \epsilon^{3\bar{b}c} \sin(\alpha) ( A_\rmR)_\mu^c 
\nonumber \\
( A'_\rmR )_\mu^{\bar{b}} & \equiv &  
\cos(\alpha) ( A_\rmR )_\mu^{\bar{b}}
+ \epsilon^{3\bar{b}c} \sin(\alpha) ( V_\rmR)_\mu^c 
\nonumber \\
(P'_\rmR)^3 & \equiv &
\cos(\alpha) (P_\rmR)^3 + \frac{i}{2} \sin(\alpha) (S_\rmR)^0
\nonumber \\
(P'_\rmR)^{\bar{b}} & \equiv & (P_\rmR)^{\bar{b}} \, , 
\eea
the Ward identities of flavour chiral
symmetry take the usual form~\cite{paper1}:
\beq \label{fch_WI_primed}
\widetilde\partial_\mu ( A'_\rmR)_\mu^b \simeq  2 m'_\rmR ( P'_\rmR)^b
\, , \quad\quad
\widetilde\partial_\mu ( V'_\rmR)_\mu^b \simeq  0 \, .
\eeq

This result reflects the existence of a linear mapping among 
renormalized correlation functions computed with action
parameters that correspond to different values of $\alpha$
and identical values of $g^2_\rmR$ and $m'_\rmR$~\cite{paper1}. 
The mapping between correlators of gauge invariant fields
with unequal space--time arguments at $\alpha = \bar{\alpha}$
and $\alpha =0$ reads\footnote{The
definitions~(\ref{primed_AVP}) are special cases of
those in eq.~(\ref{mapping_1}).}:
\bea
\langle \phi_{k {\rm R}}^{\prime \; (r)}(x) \dots \rangle_{g^2_\rmR,m'_\rmR,\bar{\alpha}}
& = & 
\langle \phi_{k {\rm R}}^{\;\; (r)}(x) \dots \rangle_{g^2_\rmR,m'_\rmR,0} \, ,
\nonumber \\ \label{mapping_1}
\phi_{k {\rm R}}^{\prime \; (r)}(x) & \equiv &
R_{kl}^{(r)}( \bar{\alpha} ) \phi_{l {\rm R}}^{\; \; (r)}(x) \, .
\eea
Here $\phi_{kR}^{(r)}$ 
denotes the $k$-th field component of the renormalized chiral multiplet $r$ and
$R^{(r)}(\omega)$ is the (formal) multiplet transformation matrix 
under quark transformations of the type~(\ref{tm_rot}).
The dots in~(\ref{mapping_1}) stand for further local fields, with those on the 
l.h.s.\ being related to those on the r.h.s.\ by the appropriate product of chiral 
transformation matrices $R$.
The relations~(\ref{mapping_1}) are regularization--independent properties and
imply the recovery of
parity, isospin and chirality in the continuum limit of LtmQCD~\cite{paper1}.
Moreover, it is useful to note that they look 
just as one would guess from simple formal arguments.
 
The full information on QCD with two mass--degenerate quark flavours can hence be
obtained by computing correlation functions with the action~(\ref{tmQCD_act_pb})
or~(\ref{tmQCD_act_ub}). It is understood that the continuum limit must be
approached at fixed values of the renormalized parameters, {\em including $\alpha$}.
In this way, the cutoff effects inherent to any determination of $\alpha$ at
finite lattice spacing have no impact on the (extrapolated) continuum limit 
results\footnote{Phenomena like a possible Aoki phase~\cite{aoki_phase,Singleton},
are immaterial for any $m'_\rmR \neq 0$, if the continuum limit of renormalized
quantities is taken at fixed renormalized parameters.} 

\subsubsection{Simplified operator mixings}

The regularization of QCD with fermionic action~(\ref{tmQCD_act_pb})
differs from Wilson's (improved) one as long as the quarks are massive.
In the massless theory, $\muq=0$ and $m_0 = \mc$, any difference
disappears, implying that the pattern of leading operator mixings
is {\em globally} the same as in the original formulation by Wilson. 

However, given a certain fermionic field, its physical interpretation
depends on how it transforms under the parity and isospin operations,
whose form is in turn dictated by the parameterization of the quark mass term. On the other
hand, the mixing properties of the fermionic field depend crucially on
its chiral orientation with respect to the Wilson term. Hence, for a
{\em specific} fermionic field with a given physical interpretation (e.g. the 
physical axial current), the mixing properties may change, even at the leading
level, with $\omega$ (or $\alpha$). For instance, eq.~(\ref{primed_AVP}) shows
that the component $(A'_\rmR)^{1}_\mu$ of the physical non-singlet axial current 
is given by $(A_\rmR)^{1}_\mu$ if $\alpha=\omega=0$ and by 
$(V_\rmR)^{2}_\mu$ if $\alpha=\omega=\pi/2$. The currents $(A_\rmR)^1_\mu$
and $(V_\rmR)^{2}_\mu$, which are defined in eq.~(\ref{chir_curr}),
involve different renormalization factors $\za$ and $\zv$, with
$\zv = 1$ if the one--point split definition of the bare
current $V^{2}_\mu$ is adopted. 

One can then adjust the ultraviolet
regularization so as to simplify as far as possible the renormalization 
properties of certain --but not all-- physical observables of interest.
E.g.\ working with $\omega \!= \!\pi/2$ entails
remarkable simplifications~\cite{paper1} in the renormalization of  
the decay constant of the charged
pseudoscalar mesons and the chiral condensate, as
$(A'_\rmR)^{1}_\mu = (V_\rmR)^{2}_\mu$ and $(S'_\rmR)^0 = 2i (P_\rmR)^3$.

\subsubsection{Hamiltonian formalism at fixed $a$}
Unimproved ($\csw=0$) lattice tmQCD admits a positive and selfadjoint
transfer matrix~\cite{paper2}, with the constraint $|8+2am_0| > 6$ and
no constraint on $a\muq$. Lattice correlation functions can hence be represented
as usual in terms of operator matrix elements with time--dependent coefficients.

From the symmetry properties of LtmQCD, see sect.~4.1, it follows
that the set of {\em lattice quantum numbers} for $\omega \neq 0$ 
is reduced as compared to that of Wilson's formulation ($\omega = 0$): 
parity and isospin are replaced by the quantum numbers $p_{\rm F}$
and $q_{\rm I}$ corresponding to the unphysical parity transformation
$P_{\rm F}$, eq.~(\ref{P_F_def}), and the unbroken isospin generator.
This implies, for instance, that in LtmQCD the physical vacuum and neutral pion
states are labelled by the same set of lattice quantum numbers and can
be interpolated by the same lattice field. Moreover,
lattice operator mixings are constrained
by $p_{\rm F}$ and $q_{\rm I}$ rather than parity and isospin quantum numbers. 
For instance, the operator $(A'_\rmR)^{1}_\mu$, which yields,
as $a\to 0$, the first component of the physical isotriplet axial current,
mixes at order $a\mu_\rmR$ with the second component of the physical isotriplet
vector current.

In the quantum mechanical analysis of tmQCD correlators ($\omega \neq 0$) at 
fixed $a$, one must hence include~\cite{paper1,DFH_latt01buda}
the contributions of matrix elements that
would vanish if parity and isospin were exact symmetries\footnote{The renormalized
counterparts of these matrix elements actually vanish like $a\mu_\rmR$, or
faster, in the continuum limit, because of the relations~(\ref{mapping_1}).}. 
The O($a$)--improvement is expected to reduce the size of these contributions.

\subsection{O($a$) improvement: the case $\alpha = \pi/2$}
The O($a$) improvement of LtmQCD~\cite{paper2,improv_tmQCD} is
most conveniently discussed in the
unphysical quark basis corresponding to eq.~(\ref{tmQCD_act_ub}).
Without loss of generality, one can assume that an infrared cutoff
is in place (thanks to some specific choice of external momenta 
or boundary conditions), so that all correlation functions admit
a Taylor expansion in powers of $\muq$ and $\mq$ around the massless
theory. Once the latter has been fully renormalized and O($a$) improved
in a mass independent scheme, the renormalized action
parameters of the massive theory are consistently~\cite{LUS1} defined
by eq.~(\ref{reno_param_1}), with $g_0^2$, $\mq$ and $\muq$ replaced by:
\bea \label{impr_bare_coupl}
\wt{g}_0^2 & = & g_0^2 ( 1 + \bg a \mq ) \, ,
\nonumber \\
\wt{m}_{\rm q} & = & \mq + \bm a\mq^2 + \tbm a\muq^2 \, ,
\nonumber \\
\wt{\mu}_{\rm q} & = & \muq ( 1 + \bmu a\mq)
\eea
and $\zg$, $\zm$ and $\zmu$ depending only on $\wt{g}_0^2$.
%
The absence of a term $\propto a\muq$ in the expression
for $\wt{g}_0^2$ reflects the fact that the partition function of
LtmQCD is even in $\muq$, see eq.~(\ref{ag_bound}) for the fermionic
determinant: hence an action counterterm of the form $a\muq {\rm tr}(FF)$ 
can not be generated.

Improved operators can be obtained by appropriate
subtraction of the operator mixings that come with powers of $a\mq$ or
$a\muq$. In absence of power divergent mixings, the construction
of multiplicatively renormalizable improved operators is rather simple, e.g.\
\bea \label{improv_op_ex}
(A_{\rm I})^{\bar{b}}_\mu & = & A^{\bar{b}}_\mu + \ca a \wt{\partial}_\mu P^{\bar{b}} +
                    a\muq \tba \epsilon^{3{\bar{b}}} V^c_\mu \, ,
\nonumber \\
(P_{\rm I})^{\bar{b}} & = & P^{\bar{b}} \, , 
\quad\quad \bar{b} =1,2 \, .
\eea
The corresponding renormalized and improved operators are obtained after
rescaling by suitable factors $\za(1 + \ba a\mq)$ and $\zp(1 + \bp a\mq)$.
It is useful to remark that  
the set of O($a$) counterterms that have been introduced is slightly
redundant~\cite{paper2}: one of them can be chosen arbitrarily, e.g. by 
setting $\tbm \equiv -1/2$.  

Adopting a formulation
with $|\wt{m}_{\rm q}| \simeq |\mq| \leq {\rm O}(a)$ and $\muq \neq 0$, i.e.
\beq \label{pi_over_2}
 |\alpha| = \pi/2 + {\rm O}(a) \, , 
\eeq
we find that $g^2_\rmR = \zg g_0^2$ and $m'_\rmR = \zmu \muq$ up to O($a^2$) corrections,
see eq.~(\ref{impr_bare_coupl}), and all $b$--type
improvement coefficients are not needed. One needs instead to know the $\wt{b}$-type
improvement coefficients, like $\tba$ in eq.~(\ref{improv_op_ex}), which are associated
to operator mixings that violate parity or isospin. This observation implies that one
can determine the $\wt{b}$-type coefficients by requiring the improved correlation 
functions to satisfy parity and isospin invariances\footnote{In principle, it is also possible
to get rid of the mixings of order $a\mu_\rmR$ without improving the operators: this
is the case if in the quantum mechanical analysis of the correlators
one can disentangle, 
e.g. by looking at the values of the effective masses, the contributions arising
from intermediate states with different parity and isospin.}. Implementing the
condition~(\ref{pi_over_2}) requires to know only $\csw$ and $\mc$ (unimproved).
Moreover, for $\alpha$ as in eq.~(\ref{pi_over_2}), eq.~(\ref{exact_PCVC})
is interpreted as the physical PCAC relation, see eq.(\ref{primed_AVP}):
O($a$) improved estimates of the mass,
$m_{\rm PS}$, and the decay constant, $F_{\rm PS}$, of the charged pseudoscalar
meson can hence be obtained~\cite{improv_tmQCD} from the correlator
\beq
\muq \wt{G}_{\rm P}(x_0) = a^3 \sum_{\bf x} \muq
\langle P^1(x) P^1(0) \rangle  \, ,
\eeq
if $\csw$ and $\mc$ are known. If one wishes to work with
$|\alpha | = \pi/2 + {\rm O}(a^2)$, an improved estimate of $\mc$
(which may require to know $\ca$) is also needed.

\section{The first non-perturbative studies}

The above theoretical understanding of LtmQCD has been checked
at the one--loop level~\cite{paper2}, as far as 
the action and the operators of eq.~(\ref{primed_AVP}) are concerned.
At the non-perturbative level, test studies have been carried out
in the quenched approximation\footnote{For all quenched
studies I conventionally set the ``physical'' scale by employing
Sommer's scale: $r_0 = 0.5$~fm.} using the non-perturbatively
improved~\cite{LSSWW} Wilson action. 
The authors of Ref.~\cite{paper3} studied LtmQCD in a box of volume $L^3 \times T$
with Schr\"odinger functional boundary conditions~\cite{paper2},
$L = 0.75$~fm, $T=2L$, $Lm'_\rmR = 0.154$, $\alpha = 1.44$
and four lattice resolutions: $a \in [0.093, 0.046]$~fm.
The scaling behaviour of some renormalized and improved
quantities (which in large volume yield the mass and the
decay constant of pseudoscalar and vector mesons) was found
to be consistent with O($a$) improvement, with the residual
cutoff effects at $a =0.093$~fm ranging from $0.5\%$ to $9\%$.

The same setup and observables have then been employed
for a study in realistically large volumes~\cite{DFH_latt01buda}: $L= 1.5$ to $2.2$~fm
and $T=(2\div3) L$, so to ensure $m_{\rm PS}L \geq 4.5$. This study was
restricted to two lattice 
resolutions, $a=0.093$ and $0.068$~fm, and, for each of them,
three sets of quark mass parameters, which correspond to $|\alpha| = \pi/2 + {\rm O}(a)$
and pseudoscalar meson masses in the range $1.85 \geq
(m_{\rm PS}/m_{\rm K^\pm})^2 \geq 0.85$. 
\begin{figure}[htb]
\begin{center}
\includegraphics[angle=-90,width=15pc]{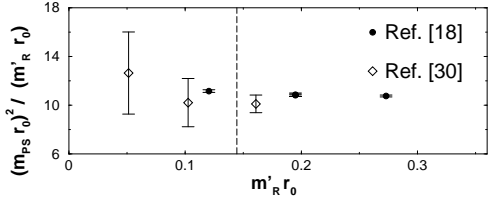}
\vspace{-25pt}
\end{center}
\caption{$r_0 \, m_{\rm PS}^2/m'_{\rm R}$ vs.\ $r_0 m'_{\rm R}$ at $\beta=6$.
The dashed vertical line corresponds to $m_{\rm PS} = m_{\rm K^\pm}$.}
\vspace{-19pt}
\label{fig_B0_beta6}
\end{figure}

Another test study of quenched LtmQCD with non-perturbatively improved action
has been performed at fixed spatial volume, $L=1.5$~fm, with $T=2L$,
lattice spacing $a=0.093$~fm, periodic boundary conditions and several
values of the quark mass~\cite{McNeile01}: the lowest value of $m_{\rm PS}$ is about
$320$~MeV, though with $m_{\rm PS}L \simeq 2.4$. In none of these studies
exceptional configurations were found, while values of $m_{\rm PS}$ 
well below $m_{\rm K^\pm}$ were reached with a moderate computational
effort: 
results for $m_{\rm PS}^2/m'_{\rm R}$ at $a=0.093$~fm from
Refs.~\cite{DFH_latt01buda,McNeile01}
are shown in Fig.~\ref{fig_B0_beta6}. 
\begin{figure}[htb]
\vspace{-10pt}
\begin{center}
\includegraphics[angle=-90,width=14pc]{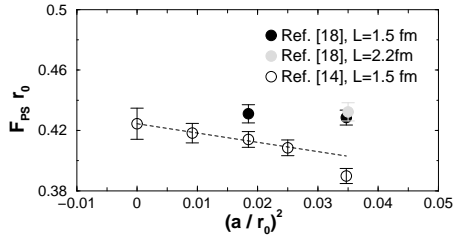}
\vspace{-30pt}
\end{center}
\caption{$F_{\rm PS}$ vs.\ $a^2$ at $m_{\rm PS} \! \simeq \! 1.2 m_{\rm K^\pm}$.
The continuum extrapolation by Ref.~\cite{garden} is also shown.}
\vspace{-10pt}
\label{fig_F_PS}
\end{figure}

\begin{figure}[htb]
\vspace{-44pt}
\begin{center}
\includegraphics[angle=0,width=15pc]{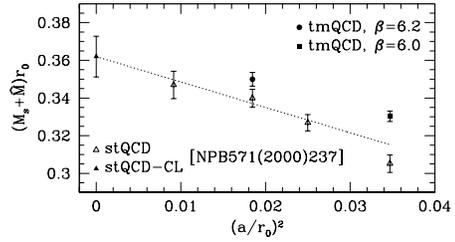}
\vspace{-79pt}
\end{center}
\caption{$(\hat{M} + M_{\rm s})\,r_0$ vs.\ $a^2/r_0^2$ for $\alpha =0$ (stQCD:~\cite{garden}) and 
$|\alpha|  \simeq  \pi/2$ (tmQCD:~\cite{DFH_latt01buda}).}
\label{fig_RGI_M}
\vspace{-10pt}
\end{figure}

The scaling behaviour of $F_{\rm PS}$ in large volume 
is presented in Fig.~\ref{fig_F_PS} for the O($a$) improved Wilson
formulations with $|\alpha| \simeq \pi/2$~\cite{DFH_latt01buda}
and $\alpha =0$~\cite{garden}. In the latter case, where four lattice
resolutions were considered to allow for continuum extrapolation, the simulation data
have been reanalysed to precisely match the renormalization conditions adopted for LtmQCD.
Following closely Ref.~\cite{garden},
an analogous comparison has been carried out, see
Fig.~\ref{fig_RGI_M}, for the combination of renormalization
group invariant quark masses $\hat{M} + M_{\rm s}$, where $\hat{M}$ is
the average mass of the $u$ and $d$ quarks. 
The results for $F_{\rm PS}\,r_0$ and $(\hat{M} + M_{\rm s})\,r_0$ 
that are obtained from the two lattice formulations
should agree in the continuum
limit: this seems to be the case within the statistical errors shown in the 
figures\footnote{For LtmQCD
$m'_{\rmR}/m_{\rm PS}^2$ at the point $m_{\rm PS} = m_{\rm K^\pm}$
could be obtained by simple {\em interpolation} of simulation data.}.
Moreover, in agreement with the indications of the scaling test at $L\simeq 0.75$~fm,
the tmQCD estimates of these quantities show pretty small cutoff effects.

\vspace{-0pt}

Standard linear solvers (CGNE, BiCGStab) were employed to invert the
(preconditioned) LtmQCD Dirac matrix. CGNE worked fine, while BiCGStab 
often got stuck for $|\alpha | \simeq |\omega | \simeq \pi/2$. The implementation of the
twisted mass term $\muq$ in unquenched algorithms (HMC, PHMC, GHMC,
multiboson) is straightforward\footnote{Unquenched simulations with
the action~(\ref{tmQCD_act_ub}) have been performed to study 
issues related to the Aoki phase~\cite{bitar}.}.

\section{$B_{\rm K}$ and the $\Delta I = 1/2$ rule}

The two-flavour formulation --LtmQCD-- that I discussed above
can be extended in a variety of ways to describe QCD with several
non-degenerate flavours of Wilson quarks. The details of the
(Wilson--like) fermionic regularization should be fixed, case by case,
so to render as simple as possible the mixing pattern of the operators that
are relevant for the physical problem of interest, while avoiding
the occurrence of spurious quark zero modes. The general reasons  
underlying this possibility are discussed in sect.$\;4.2.1$.

For the computation of $B_{\rm K}$ ($K^0$--$\bar{K}^0$ mixing)
it is convenient to adopt the lattice formulation with fermionic
action density:
\beq
{\cal L}_{\rm F} = \qbar\! \left(
D_{\rm W}^c + i m'_{\rm q\,l} \gamma_5 \tau^3
\right)\! q + 
\bar{s}\! \left( D_{\rm W}^c + m'_{\rm q\,s} \right)\! s   
\eeq
which corresponds to take $\alpha = \pi/2$ for the two degenerate light 
flavours and $\alpha =0$ for the strange quark. In this quark basis the 
operator for (physical) parity--even $\Delta S =2$ transitions reads~\cite{paper1}
\beq
 O^{\prime (\Delta S =2)}_{VV+AA} 
= - 2i (\bar{s} \gamma_\mu d)
  (\bar{s} \gamma_\mu \gamma_5 d) \equiv -2i V_{sd} A_{sd}
\eeq
and is multiplicatively renormalizable. A computation of $B_{\rm K}$
based on this approach is currently in progress~\cite{BK_SF_tm}. 

For the computation of CP conserving $\Delta S =1$ matrix elements of
the weak effective Hamiltonian with active charm flavour, it is convenient 
to adopt the following lattice formulation~\cite{DS_eq_1_tm}:
\beq
{\cal L}_{\rm F} = \psibar \left(
D_{\rm W}^c + \widehat{m}_{\rm q} + i\widehat{\mu}_{\rm q} \gamma_5 \tau^3
\right) \psi \,
\eeq
with $\psi^{\rm T}\!=(u,d;s,c)$ and quark mass matrices
\bea
\widehat{m}_{\rm q} & = & {\rm diag}(m_{\rm q\,l}, m_{\rm q\,l}; m_{\rm q\,s}, m_{\rm q\,c}) 
\nonumber \\
\widehat{\mu}_{\rm q} & = & {\rm diag}(\mu_{\rm q\,l}, -\mu_{\rm q\,l}; 
                                       \mu_{\rm q\,s}, \mu_{\rm q\,c})
\, . 
\eea
Thinking of $\psi$ as a pair of doublets and setting 
\beq
{ \mu_{\rm q\,l} \over m_{\rm q\,l} } \equiv \tan \omega_l \, , \quad\quad
{ \mu_{\rm q\,s} \over m_{\rm q\,s} } = - { \mu_{\rm q\,c} \over m_{\rm q\,c} } \equiv
\tan \omega_h \, ,
\eeq
one can show~\cite{DS_eq_1_tm} --by arguments analogous to those of sect.$\;4.2$-- that the operator
$$ O_{VA+AV}^\pm = [ (V_{sd}A_{uu} + A_{sd}V_{uu}
 \pm \dots ) - (u \to c) ] $$
must be interpreted as the parity--even operator $i\,O^{\prime \;\; \pm }_{VV+AA}$
(relevant for $K \to \pi$ transitions), if $\omega_l = \omega_h = \pi/2$,
and as the parity--odd operator $O^{\prime \;\; \pm }_{VA+AV}$
(relevant for $K \to \pi\pi$ transitions), if $\omega_l = -\omega_h = \pi/2$.
Symmetries and power counting imply that $O_{VA+AV}^\pm$ does not mix with dimension
six operators, while the leading mixing with dimension three operators has a 
coefficient  
$\propto \, a^{-1} (\mu_{\rm q\,c} - \mu_{\rm q\,l})(\mu_{\rm q\,s} - \mu_{\rm q\,l})$.
Quadratic divergences in $K \to \pi$ matrix elements and spurious quark
zero modes in general are hence avoided. 
 
\section{Conclusions}
Wilson fermions with chirally twisted mass can provide a variety
of alternative regularizations for lattice QCD. Thanks to an up to now
unexploited freedom in formulating lattice QCD with Wilson quarks, they
allow to avoid some of the most serious problems due to lack of
lattice chiral symmetry, while preserving ultralocality of
the action and hence a moderate computational cost.

\section*{Acknowledgements}
I would like to thank M.~Della~Morte, P.A.~Grassi, J.~Heitger,
C.~Pena, S.~Sint, A.~Vladikas and P.~Weisz 
for the pleasant and stimulating collaboration work. 
I am also indebted with M.~Della~Morte and S.~Sint for 
critical remarks on this write-up and
with M.~L\"uscher, G.C.~Rossi and R.~Sommer
for several, very useful discussions.

\end{document}